\documentclass[preprint]{revtex4}

\usepackage[dvips]{graphicx}
\usepackage{multirow}
\usepackage[dvips]{color}

\newcommand{\aprime}{\ensuremath{\alpha^{\prime}}}
\newcommand{\dprime}{\ensuremath{\delta^{\prime}}}

\begin{document}

\title{Composition Profiles within Al$_3$Li and Al$_3$Sc/Al$_3$Li Nanoscale Precipitates in Aluminum}

\author{Matthew E. Krug}
\author{David C. Dunand}
\author{David N. Seidman}
\email{d-seidman@northwestern.edu}
\affiliation{
Department of Materials Science and Engineering and the Northwestern University Center for Atom-Probe Tomography (NUCAPT)\\
Evanston, IL 60208--3108, USA}
\homepage{http://arc.nucapt.northwestern.edu/}
\date{January 28, 2008}

\begin{abstract}
An Al--11.3Li--0.11Sc (at.\ \%) alloy was double-aged to induce first \aprime-Al$_3$Sc and then \dprime-Al$_3$Li precipitates.  Atom-probe tomography  revealed both single-phase \dprime-precipitates and core-shell \aprime/\dprime-precipitates (with respective average radii of 16 and 27 nm, and respective volume fractions of 12 and 9\%) conferring a high strength to the alloy.  Although the \dprime-shells contain little Sc ($\sim$0.027~at.~\%), the \aprime-cores have a high Li content, with an average composition of Al$_{0.72}$(Sc$_{0.17}$Li$_{0.11}$).  The Li concentrations within the \dprime-phase and the Li interfacial excess at the \dprime/\aprime-interface both exhibit wide precipitate-to-precipitate variations.
\end{abstract}
\maketitle

Each weight percent Li added to Al alloys results in a decrease in density and an increase in Young's modulus of nearly 3 and 6 percent, respectively, leading to improvements in specific stiffness that result in weight savings for structural applications~\cite{Westwood_1990}.  Additionally a large strength increment is produced by the formation of L1$_2$-ordered, coherent, \dprime-Al$_3$Li precipitates~\cite{Sanders_1980}, with an achievable volume fraction of up to $\sim$30\% due to the high solubility of Li in Al~\cite{McAlister_1982}.  Hypo-eutectic Al--Sc alloys display a high density of L1$_2$-ordered, coherent, \aprime-Al$_3$Sc precipitates which, when compared to the \dprime-Al$_3$Li precipitates, have the following differences: (i) they are stable, rather than metastable; (ii) they have much better coarsening resistance due to the smaller diffusivity of Sc as compared to Li in Al~\cite{Minamino_etal_1987, Fujikawa_1997}; and (iii) they exhibit a much smaller volume fraction ($\sim$0.93\%) because of the limited solid-solubility of Sc in Al~\cite{Murray_1998}.  Moreover, Sc additions confer greater coarsening resistance to the \dprime-precipitates in Al--Li alloys~\cite{Miura_etal_1994, Joh_Yamada_1999, Beresina_etal_2002} due to the attractive interaction between Sc atoms and vacancies~\cite{Hashimoto_etal_1979}. Hence, Al-based alloys with both Li and Sc additions offer the potential for technologically interesting alloys with high specific stiffness and strength (from Li) and improved thermal stability and creep resistance (from Sc).

Several TEM studies conducted on Al--Li--Sc alloys revealed the development of both single-phase \dprime-precipitates as well as complex precipitates consisting of an \aprime-core surrounded by a \dprime-shell~\cite{Miura_etal_1994, BEREZINAAL_etal_1991, Beresina_etal_2002}.  It was found that for an Al--8.8Li--0.11Sc (at.\ \%) alloy the maximum strengthening effect was attained for a duplex heat treatment, wherein \aprime-precipitation was first induced at 400~$^{\circ}$C, followed by \dprime-precipitation at 200~$^{\circ}$C~\cite{Miura_etal_1994}.  In the present study, we use atom-probe tomography (APT)~\cite{Seidman_2007, Kelly_etal_2007} to characterize at the atomic level the \aprime- and \dprime-phases in an Al--Li--Sc alloy, allowing precise determination of their volumetric and interfacial chemical compositions, which were not accessible in the prior TEM studies~\cite{Miura_etal_1994, BEREZINAAL_etal_1991, Beresina_etal_2002}, but which were examined in a recently published APT/TEM study of a quaternary Al--6.30Li--0.36Sc--0.13Zr (at.\ \%) alloy~\cite{Radmilovic_etal_2008}.

An Al--11.3Li--0.11Sc alloy (at.\ \%), was cast from a 99.9\% pure Al--1.3Sc (at.\ \%) master alloy, 99.9\% pure Li and 99.999\% pure Al.  To prevent oxidation and achieve rapid heat transfer, all heat treatments were performed in molten salt (NaNO$_2$--NaNO$_3$--KNO$_3$) baths on 10$\times$10$\times$10 mm$^3$ specimens, with temperature controlled to $\pm$2~$^{\circ}$C.  After homogenizing for 24 h at 598~$^{\circ}$C, the specimens were quenched directly to 350~$^{\circ}$C and held for 6 h to induce \aprime-precipitation. Next they were held for 2 min at 510~$^{\circ}$C to dissolve any stable AlLi which may have formed at grain boundaries during the prior step~\cite{Prasad_etal_2003}, and then quenched to 200~$^{\circ}$C where they were aged for 24 h to induce \dprime-precipitation. Because \dprime-precipitation occurs even during rapid quenching in alloys richer than $\sim$6~at.~\% Li~\cite{Williams_1980}, specimens to be examined by APT were transferred from each heat treating temperature directly to the subsequent temperature, with no intermediate quench.  The specimens were finally brine-quenched to -15~$^{\circ}$C and transferred to liquid nitrogen (78 K) within 30 min to minimize diffusion of Li.  All specimens were removed from material near the center of the cube, away from the Li-depleted surface.

After each of the heat treatments, a specimen was brine-quenched to monitor the effect of each step on mechanical strength, as measured by Vickers microhardness tests.  A specimen subjected to the complete heat treatment was prepared into a sharply pointed tip by electropolishing in a solution of 10\% perchloric acid in acetic acid, followed by a solution of 2\% perchloric acid in butoxyethanol.  The tip was analyzed with a local-electrode atom-probe (LEAP\textsuperscript{\texttrademark}-3000X Si, Imago Scientific Instruments) at 30 K, using voltage pulsing with a 200 kHz repetition rate, and a 15\% pulse fraction (pulse voltage/steady-state dc voltage). \textsc{ivas}$^{\textregistered}$ (Imago) was used to analyze quantitatively the data.

The high baseline microhardness of the homogenized and quenched alloy (620 MPa) is indicative of \dprime-precipitation.  The first 350~$^{\circ}$C aging treatment increases hardness to 790 MPa due to \aprime-nucleation and growth, and a further increase in hardness to 1330 MPa occurs after the second 200~$^{\circ}$C aging step due to the \dprime-precipitates.  The latter value is similar to a peak hardness obtained in a prior study of $\sim$1275 MPa~\cite{Miura_etal_1994}, wherein a comparable aging treatment was used for an alloy with the same Sc concentration and a Li concentration of 8.8~at.~\%.

\begin{figure}
  \begin{center}
    \includegraphics[width=15cm]{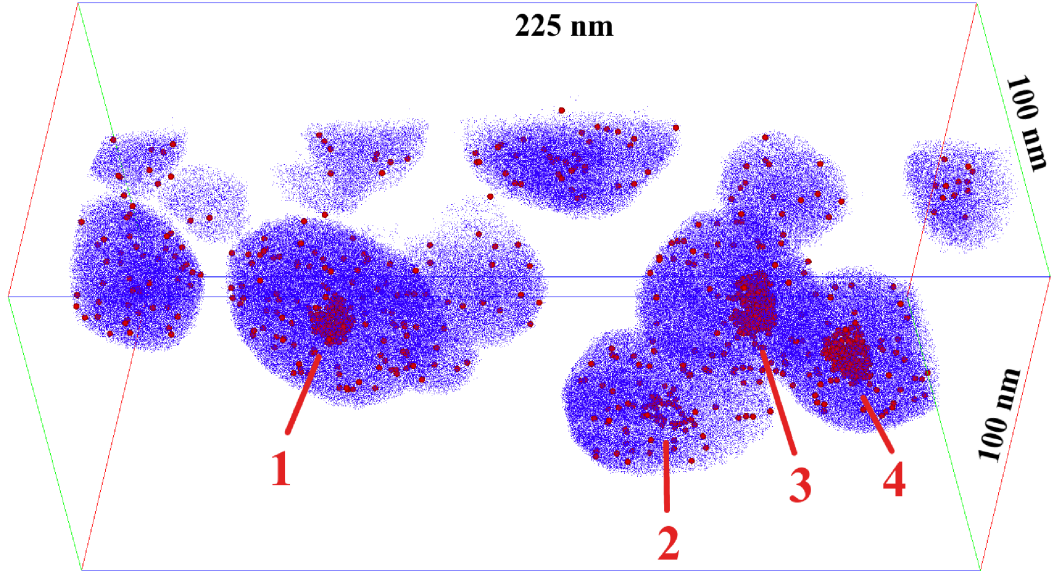}
  \end{center}
    \caption{A LEAP\textsuperscript{\texttrademark} tomographic reconstruction of a doubly-aged specimen showing 11 single-phase \dprime-precipitates and four numbered core-shell \dprime/\aprime-precipitates (only a small portion of the \aprime-core of precipitate number two is contained in the reconstructed volume).  Lithium atoms are in blue, Sc atoms are in red, and Al atoms are omitted for clarity, as are all $\alpha$-matrix atoms; the atomic diameters are not to scale.}\label{fig:tip}
\end{figure}

Fig.~\ref{fig:tip} displays a 50$\times$10$^6$ atom partial reconstruction of a specimen that had completed the duplex aging treatment to promote first \aprime- and then \dprime-precipitates.  Two types of precipitates are present: single-phase \dprime-precipitates with an average radius \ensuremath{\left\langle{R}\right\rangle}=16 nm, and dual-phase precipitates (\ensuremath{\left\langle{R}\right\rangle}=27 nm; indicated by red numbers in Fig.~\ref{fig:tip}) with a \dprime-shell surrounding an \aprime-core (\ensuremath{\left\langle{R}\right\rangle}$_{core}$=6 nm).  The disparate sizes of these two types of precipitates is in agreement with previous TEM studies on Al--Li--Zr, an analogous system~\cite{Aydinol_Bor_1994, Gu_etal_1985}.  With continued aging, it is anticipated that the core-shell precipitates, where the \dprime-shell probably nucleated heterogeneously on the preexisting \aprime-precipitates, will grow at the expense of the single-phase \dprime-precipitates~\cite{BEREZINAAL_etal_1991, Dlubek_etal_1992}.  The  number density of all precipitates is 1$\pm$0.3$\times10^{22}$~$m^{-3}$, with a volume fraction of 21\%.  Of the 46 precipitates with centroids within the 125$\times$10$^6$ atom full reconstruction, six (13\% by number, 42\% by volume) are the larger core-shell type and the remainder are the smaller, single-phase \dprime-precipitates.

\begin{figure}
  \begin{center}
    \includegraphics[width=15cm]{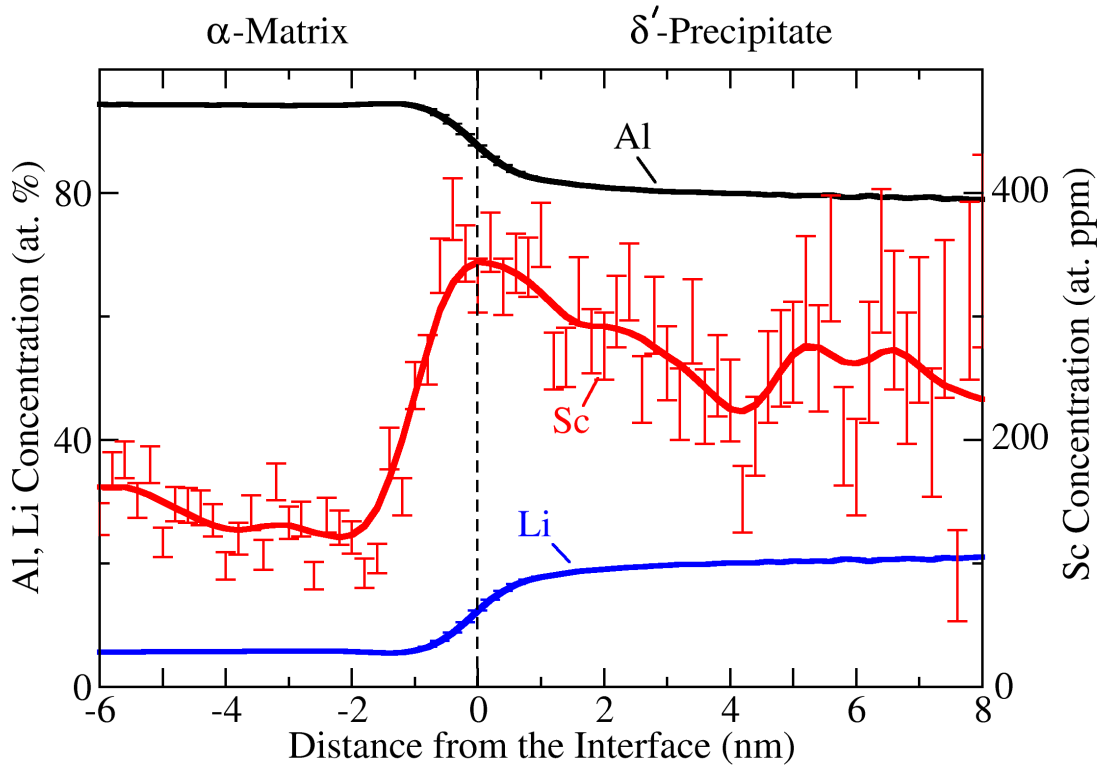}
  \end{center}
    \caption{Proximity histogram of 40 core-free \dprime-precipitates showing the average concentration of Al, Li, and Sc as a function of distance from the $\alpha$-matrix/\dprime-interface, as defined by a 12.5~at.~\% Li isoconcentration surface.}\label{fig:cfpprox}
\end{figure}

\begin{figure}[h!]
  \begin{center}
    \includegraphics[width=15cm]{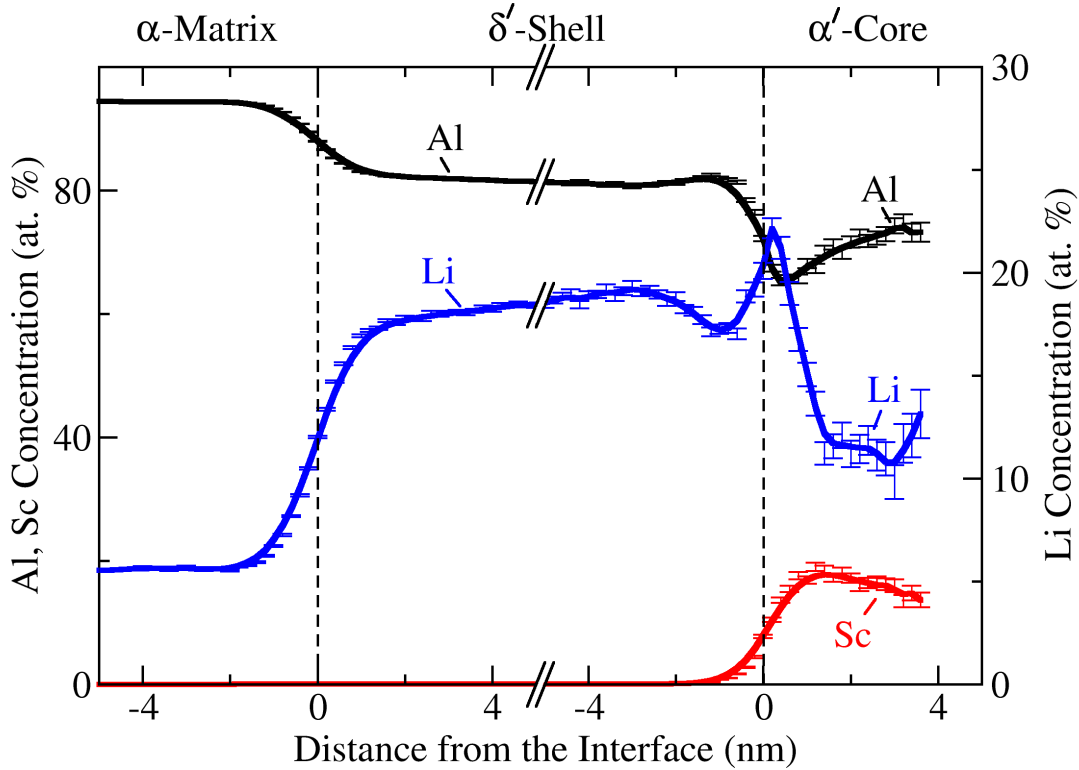}
  \end{center}
  \vspace{-20pt}  
  \caption{Combined proximity histograms of six core-shell precipitates displaying the average concentration of Al, Li, and Sc as a function of distance from the $\alpha$-matrix/\dprime-interface, and from the \dprime/\aprime-interface, as defined by a 12.5~at.~\% Li isoconcentration surface and a 9~at.~\% Sc isoconcentration surface, respectively.  The interruption of the $x$-axis in the \dprime-shell removes, on average, 11 nm from the proxigram in a region where the concentration profiles are flat.}\label{fig:cspprox}
\end{figure}

Table~\ref{tab:table1} presents the average compositions of the different phases present in the $\alpha$-matrix and in both types of precipitates after the full heat-treatment, as measured by APT by constructing proximity histograms (proxigrams) at the different heterophase interfaces and determining the composition of the two adjoining phases away from the interface, within the region where the concentration profiles are flat.  The $\alpha$-matrix concentration of 5.7~at.~\% Li is in approximate agreement with the value predicted for the binary Al--Li system at 200~$^{\circ}$C (6.3~at.~\%)~\cite{Noble_Bray_1998}.  At 141~at.~ppm, the Sc concentration of the $\alpha$-matrix is about five times greater than that predicted by extrapolation to 350~$^{\circ}$C of the \aprime-solvus in the binary Al-Sc system~\cite{Murray_1998}.  Although Li may increase the solid solubility of Sc in Al, it is also possible that the system did not achieve equilibrium during the aging treatments at 350~$^{\circ}$C.  The average Li concentration is 19.8~at.~\% in the smaller single-phase \dprime-precipitates and 18.8~at.~\% in the \dprime-shell of the larger core-shell precipitates.  This difference is probably not significant, however, since in both types of precipitates the Li concentration in \dprime\ varies strongly from precipitate-to-precipitate: in the six core-shell precipitates the Li concentration in the \dprime-phase varies between $\sim$12--21~at.~\%, whereas in a random selection of six single-phase precipitates the Li concentration in the \dprime-phase varies between $\sim$16--21~at.~\%.  In neither type of precipitate is any correlation between the radius and Li concentration apparent.  The average Li concentrations in the \dprime-phase of both types of precipitates are at the lower end of the wide range reported for the binary Al--Li system, (19--23~at.~\%)~\cite{Hallstedt_Kim_2007}, indicating that nanosize \dprime-trialuminides are strongly sub-stoichiometric in Li.  A similar substoichiometry in Li was measured in the \dprime-shell of a core-shell precipitate in Ref.~\cite{Radmilovic_etal_2008}, in which precipitate-to-precipitate variability in Li concentration was not explored due to the smaller number of precipitates examined by APT.  Here, the Sc concentration in the \dprime-phase is the same, within measurement error, in both types of precipitates ($\sim$270~at.~ppm) and is about double that of the surrounding $\alpha$-matrix ($\sim$140~at.~ppm).  This partitioning of Sc to the \dprime-phase is clearly demonstrated in Fig.~\ref{fig:cfpprox} in a proxigram aggregated over 40 precipitates, showing the concentration of Sc as a function of radial distance from the $\alpha$/\dprime-interface in single-phase \dprime-precipitates.  Concentration error bars in Figs.~\ref{fig:cfpprox}~and~\ref{fig:cspprox} and Table~\ref{tab:table1} are calculated as $\sqrt{c_i(1-c_i)/N_{total}}$, where $c_i$ is the atomic fraction of element $i$ and $N_{total}$ is the total number of atoms detected.  The proxigrams and table do not convey information about precipitate-to-precipitate variability, which is significant, as discussed above.

\begin{table}
\caption{\label{tab:table1}Composition of constituent phases determined by LEAP\textsuperscript{\texttrademark} tomographic analysis (overall macroscopic composition measured by directly coupled plasma mass spectroscopy given for comparison).}
\begin{ruledtabular}
\begin{tabular}{lccc}
& Al (at.\ \%)& Li (at.\ \%)& Sc (at.\ \%)\\
\hline
\hline
Macroscopic Specimen & 89$\pm$1 & 11$\pm$1 & 0.11$\pm$0.003\\
\hline
$\alpha$-Matrix & 94.29$\pm$0.005 & 5.70$\pm$0.005 & 0.014$\pm$0.0003\\
\hline
Core-Free \dprime-Precipitates & 80.2$\pm$0.01 & 19.8$\pm$0.01 & 0.027$\pm$0.0006 \\
\hline
\dprime-Precipitate Shells & 81.12$\pm$0.005 & 18.83$\pm$0.005 & 0.026$\pm$0.0008 \\
\hline
\aprime-Precipitate Cores & 71.8$\pm$0.03 & 11.5$\pm$0.02 & 16.7$\pm$0.02 \\
\end{tabular}
\end{ruledtabular}
\end{table}

A proxigram showing the Al, Li, and Sc concentrations as a function of distance from the $\alpha$-matrix/\dprime- and the \dprime/\aprime-heterophase interfaces is presented in Fig.~\ref{fig:cspprox}.  It is constructed from the four core-shell precipitates visible in Fig.~\ref{fig:tip}, as well as two core-shell precipitates from a second data set from the same tip.  Because a small number of core-shell precipitates are present in the analyzed volume, a single precipitate can significantly influence the average proxigram.  For instance, traversing the \dprime-shell from the $\alpha$-matrix toward the \aprime-core, the average Li concentration increases from $\sim$17.5 to 19.5~at.~\% in Fig.~\ref{fig:cspprox}.  Of the six precipitates averaged to construct this proxigram, the Li concentration in the shell was, however, invariant as a function of radial position to within measurement error in all but one.  Also, a large confined Li interfacial excess is apparent at the \dprime/\aprime-interface in the average proxigram of Fig.~\ref{fig:cspprox}.  A confined interfacial excess of Li, though less substantial than the average value shown in Fig.~\ref{fig:cspprox}, may be discerned at this heterophase interface in three of the six core-shell precipitates.  The magnitude of the average Li interfacial excess is, however, dominated by a fourth precipitate in which the Li concentration at this heterophase interface reached 34~at.~\%.  The cores of the core-shell precipitates are characterized by a strong partitioning of Sc, consistent with the thermal history of the specimen, which promoted precipitation first of \aprime\ followed by \dprime.  Lithium is also present in the core, indicating that it has high solubility in the \aprime-phase.  While this is in qualitative agreement with Ref.~\cite{Radmilovic_etal_2008}, a quantitative comparison is complicated by the Zr-addition, the smaller Li concentration, the larger Sc concentration (hyper-eutectic), and the different aging conditions employed in that study.  Table~\ref{tab:table1} demonstrates that the composition of the inner portion of the Sc-rich cores (i.e., excluding the interfacial excess of Li) is close to Al$_3$(Sc$_{1-x}$Li$_x$) with $x$ = 0.4, which suggests that in Al$_3$Sc (L1$_2$), Li substitutes on the Sc sublattice.  The confined Li interfacial excess and corresponding depletion of Al at the core-shell heterophase interface suggest, however, that Li can also substitute on the Al sublattice.  This indicates that the kinetic pathway for the temporal evolution of these precipitates is complex, which is only evident with detailed APT experiments and analyses.  It is also possible, however, that the 510~$^{\circ}$C heat treatment between the 350~$^{\circ}$C and 200~$^{\circ}$C heat treatment steps (to dissolve AlLi and to form \aprime- then \dprime-precipitates, respectively) plays a role in determining the concentration profiles at this heterophase interface.

In summary, a detailed investigation by APT of the chemical composition of precipitates in the Al--Li--Sc system has been conducted.  A duplex heat treatment was utilized to promote sequential formation of \aprime-Al$_3$Sc and \dprime-Al$_3$Li precipitates. Two populations of precipitates are observed: larger core-shell \aprime/\dprime-precipitates and smaller single-phase \dprime-precipitates.  Investigation by APT indicates that: (i) the Li concentration in the \dprime-phase of both core-shell and single-phase precipitates is highly variable (12--21~at.~\%); (ii) a sizeable confined (non-monotonic) Li interfacial excess is present at the \dprime/\aprime-interface, which also varies widely from precipitate-to-precipitate; and (iii) the Li concentration of the \aprime-cores, with an average composition of Al$_{0.72}$(Sc$_{0.17}$Li$_{0.11}$), is high, indicating that Al$_3$Sc has a high solubility for Li.

\begin{acknowledgments}
This research is supported by the United States Department of Energy through grant DE--FG02--98ER45721.  The LEAP tomograph was purchased with funding from the NSF--MRI (Grant DMR--0420532) and ONR--DURIP (Grant N00014--0400798) programs.
\end{acknowledgments}

\end{document}